\documentclass[prc,amsmath,twocolumn,showkeys,showpacs,superscriptaddress]{revtex4}
\usepackage{gensymb}
\usepackage{graphicx,color}
\usepackage{amssymb}
\usepackage{enumerate}
\usepackage{verbatim}
\usepackage{natbib}

\begin{document}
  \newcommand {\nc} {\newcommand}
  \nc {\Sec} [1] {Sec.~\ref{#1}}
  \nc {\IR} [1] {\textcolor{red}{#1}} 
  \nc {\IB} [1] {\textcolor{blue}{#1}} 

\title{Energy dependence of non-local optical potentials}

\author{A.~E.~Lovell}
\affiliation{National Superconducting Cyclotron Laboratory, Michigan State University, East Lansing, MI 48824}
\affiliation{Department of Physics and Astronomy, Michigan State University, East Lansing, MI 48824-1321}
\author{P.-L.~Bacq}
\affiliation{Physique Nucl\' eaire et Physique Quantique, Universit\'e Libre de Bruxelles (ULB), B-1050 Brussels}
\affiliation{National Superconducting Cyclotron Laboratory, Michigan State University, East Lansing, MI 48824}
\affiliation{Department of Physics and Astronomy, Michigan State University, East Lansing, MI 48824-1321}
\author{P.~Capel}
\affiliation{Physique Nucl\' eaire et Physique Quantique, Universit\'e Libre de Bruxelles (ULB), B-1050 Brussels}
\affiliation{Institut f\"ur Kernphysik, Technische Universit\"at Darmstadt, 64289 Darmstadt, Germany}
\author{F.~M.~Nunes}
\email{nunes@nscl.msu.edu}
\affiliation{National Superconducting Cyclotron Laboratory, Michigan State University, East Lansing, MI 48824}
\affiliation{Department of Physics and Astronomy, Michigan State University, East Lansing, MI 48824-1321}
\author{L.~J.~Titus}
\affiliation{National Superconducting Cyclotron Laboratory, Michigan State University, East Lansing, MI 48824}
\affiliation{Department of Physics and Astronomy, Michigan State University, East Lansing, MI 48824-1321}

\date{\today}


\begin{abstract}
Recently a variety of studies have shown the importance of including non-locality in the description of reactions.
The goal of this work is to revisit the phenomenological approach to determining non-local optical potentials from elastic scattering.  
We perform a  $\chi^2$ analysis of neutron elastic scattering data off $^{40}$Ca, $^{90}$Zr and $^{208}$Pb at energies $E\approx 5-40$ MeV, assuming a Perey and Buck \cite{Perey_np1962} or Tian, Pang, and Ma \cite{Tian_ijmpe2015} non-local form for the optical potential.
We introduce 
energy and asymmetry dependencies in the imaginary part of the potential
and refit the data to obtain a global parameterization.
Independently of the starting point in the minimization procedure, an energy dependence in the imaginary  depth is required for a good description of the data across the included energy range. We present two parameterizations, both of which represent an improvement over the original potentials for the fitted nuclei as well as for other nuclei not included in our fit. 
Our results  show that, even when including the standard Gaussian non-locality in optical potentials, a significant energy dependence is required to describe elastic-scattering data. 
\end{abstract}


\maketitle

{\it Introduction}: 
Optical potentials are commonly used in nuclear-reaction theory as effective interactions that take into account the complexity of the many-body effects in nucleon-nucleus scattering. These effective potentials are often determined from fitting to data, mostly elastic-scattering angular distributions.
The vast majority of these global fits have assumed the interaction is local and strongly energy dependent. However, it is well understood from many-body structure, that such effective interactions are intrinsically non-local. Even at the mean field level,  the exchange term in Hartree-Fock theory, originating from antisymmetrization, introduces an explicit non-local potential \cite{Dickhoff_Book}. 
In addition, coupling from the elastic channel to all other channels not explicitly included in the model space also gives rise to non-locality \cite{Feshbach_ap1958, Feshbach_ap1962,Fraser_epja2008, Rawitscher_prc1994, Rawitscher_npa1987}.

No matter how complex the target, it is always possible to design a local optical potential that fits elastic scattering at a given energy, as long as enough degrees of freedom are included in the parameterization. To illustrate this fact, we note the work on neutron scattering on one of the most challenging targets,  $^9$Be \cite{Bonaccorso_prc2014}. The strong clusterization in this system produces large coupling effects (non-local by nature) that can be mimicked by local strong surface terms, both in the real and imaginary parts of the optical potential. The resulting energy dependence is significant and intricate. However, despite it being convenient to assume that all non-local effects can be encapsulated in the energy dependence of local optical potentials, in the last few years numerous studies have demonstrated the importance of including non-locality explicitly in the predictions of reaction observables \cite{Titus_prc2014,Ross_prc2015,Titus_prc2016,Ross_prc2016,Timofeyuk_prc2013,Waldecker_prc2016,Bailey_prl2016}.
These studies have shown that non-locality in optical potentials can greatly affect the transfer cross sections in both shape and magnitude.
For this reason it is important to revisit the issue of non-locality in optical potentials and find ways to constrain it. This study is one step along the way and addresses the question: does nucleon-target elastic-scattering data call for an energy dependence in the optical potential, when  non-locality is included explicitly? 

Over the last few decades there have been only two studies aimed at providing an explicitly non-local phenomenological nucleon-target potential, namely the work in the sixties by Perey and Buck (PB) \cite{Perey_np1962} and the more recent effort by Tian, Pang and Ma (TPM) \cite{Tian_ijmpe2015}.
In the first case, only two data sets were used: the angular distribution for the elastic scattering of $n+^{208}$Pb at $7.0$ and $14.5$ MeV. The PB potential consists of a non-local real volume term, a non-local imaginary surface term, and a local real spin-orbit potential. 
The non-locality is of Gaussian form with an ad-hoc range $\beta=0.85$ fm. The TPM for neutrons was fitted to elastic-scattering data on $^{32}$S, $^{56}$Fe, $^{120}$Sn and $^{208}$Pb for energies in the range of $E=8$--$30$~MeV.  
In addition to the non-local real volume term and the non-local imaginary surface term, the TPM potential includes a non-local imaginary volume term.  As for PB, the non-locality is also assumed to be Gaussian but the range  $\beta$ is an additional parameter in the fit.
In both cases, the  fitted parameters were assumed to be energy and mass {\it independent}.

Although the potential itself is not an observable, microscopic theories should be able to provide insight into the issue of non-locality and energy dependence. There have been many efforts to derive optical potentials from microscopic theories. The link between the self-energy and the optical potential, explored by Mahaux and Sartor \cite{Mahaux_anp1991} was implemented for a number of targets, and is known as the dispersive optical model (DOM)\cite{Dickhoff_prc2010, Mahzoon_prl2014}.
There are also ongoing efforts of extracting the optical potential from ab initio theories. In Ref.~\cite{Rotureau_prc2017}, the optical potential is extracted from the many-body Green's function generated in a coupled-cluster calculation. The resulting effective potential is strongly non-local and energy dependent.
In these studies, the non-locality produced is not described by a simple Gaussian shape and the range is larger than assumed in PB \cite{Mahzoon_prl2014,Ross_prc2015, Rotureau_prc2017}.


At present, non-local nucleon optical potentials extracted from state-of-the-art ab initio theories are still unable to provide a detailed description of the data \cite{Rotureau_prc2017}.
Moreover they are difficult to implement within direct-reaction models.
It is therefore important to revisit the phenomenological approach.
The goal of the current study is to investigate, once a standard Gaussian non-locality is introduced explicitly, whether elastic-scattering data requires an energy dependence in the potential and explore the possible parameterization of that dependence.
We take as  starting points, the PB and TPM parameterizations. We consider three spherical targets for which there is good quality neutron scattering data  ($^{40}$Ca, $^{90}$Zr and $^{208}$Pb) and a range of neutron energies $5\leq E \leq 40$ MeV.  To obtain a practical phenomenological global non-local potential we use a single-channel approach. Although coupling to other channels can produce dynamic polarization in the scattering process, these effects are averaged out in the optical model, especially for spherical targets.
Consequently, nucleon elastic scattering off $^{40}$Ca, $^{90}$Zr and $^{208}$Pb is usually well described using simple local optical potentials.
Note that neutron scattering on light nuclei exhibiting strong clusterization features are extremely challenging to describe in a global approach, and therefore fall outside the scope of this work. Dedicated studies for those systems, such as Ref.~\cite{Bonaccorso_prc2014}, are necessary.

We analyze the data with PB-like and TPM-like interactions,  considering the issue of the dependence on energy and asymmetry $\frac{N-Z}{A}$, where $N$, $Z$ and $A$ are, respectively, the neutron, proton and total nucleon numbers of the target.
Our results demonstrate unequivocally that the data calls for an energy dependence in the imaginary part of the optical potential.
To test the predictive power of the obtained parameterizations, we apply them to other cases ($^{27}$Al and $^{118}$Sn).
Finally, we discuss the limitations of the present construction.\\

{\it PB and TPM parameterizations:}
In this work, we have coupled the code NLAT \cite{nlat} to sfresco \cite{fresco} in order to perform a $\chi^2$ minimization of angular distributions produced in the optical model with non-local potentials. We have selected 24 sets of data for neutron elastic scattering on three different  targets at several beam energies: $^{40}$Ca 
($E=9.9$, 11.9, 13.9, 16.9, 21.7, 25.5, 30.1, 40.1 MeV), $^{90}$Zr ($E=5.9$, 7.0, 8.0, 10.0, 11.0, 24.0 MeV),  $^{208}$Pb ($E=7.0$, 9.0, 11.0, 14.6, 16.9, 20.0, 22.0, 26.0, 30.3, 40.0 MeV) \cite{40Ca11data,40Ca913data,40Ca16data,40Ca21253040data,90Zr56data,90Zr81024data,90Zr11data,208Pb791126data,208Pb14data,208Pb16data,208Pb2022data,208Pb3040data}. We made sure all targets were spherical nuclei, spanning a wide range of asymmetry, 
and that the data covered most of the angular range.
Although most often systematic errors on elastic-scattering data are not discussed in the publications, we assume these dominate the uncertainties in the cross sections and take a 10\% error for all data points to account for these. 

We first evaluate the $\chi^2$ obtained when using either the original PB or the original TPM potential, for each data set we consider. Because PB only fits elastic scattering at $E<15$ MeV, it does worse at higher energies. TMP on the other hand does better than PB in the range $20<E<30$ MeV but worse at lower energies. At $E=40$ MeV, TPM does poorly over the whole angular range just like PB, suggesting that extrapolations beyond the fitted range are not reliable.

In columns 3 and 4 of Table~\ref{tab:chi2}, we show the $\chi^2$ compiled by summing the various sets over two energy bins.
Because we expect the optical model approach to work poorly at backward angles, we have considered  
$\chi_{\theta <100}^2$, restricting the angles to $\theta<100^\circ$. By comparing columns 3 and 4, one can verify that there are significant differences between
$\chi_{\theta <100}^2$ and the full $\chi_{\rm tot}^2$, calculated with the whole angular range.
As pointed out above, PB does better at lower beam energies, while TPM overall provides a better description at the higher beam energies. The main difference of TPM compared to PB, is the inclusion of the volume imaginary term which is expected to be necessary in describing data at the higher energies.

\begin{table}[t!]
\centering
\begin{tabular}{| c | c | r | r | r |r | r | r |}
\hline
 & $E$ (MeV) & $\chi_{\theta <100}^2$ & $\chi_{\rm tot}^2$   &  $\chi_{\theta <100}^2(E)$ & $\chi_{\rm tot}^2(E)$  \\
\hline 
PB & $<20$		& 116 &  136 & 92  & 121   \\ 
PB &  $\geq 20$ 	& 640 & 465 &  61 &  136  \\ 
\hline
TPM & $<20$ 	& 131 & 230 &  109 &  158 \\ 
TPM & $\geq 20$	& 158 & 182 & 82  & 177  \\ 
\hline
\end{tabular}
\caption{Summed $\chi^2$ for the various reactions here considered: the original parameterization is shown in the first column, the energy range for the neutron in the laboratory is the second column, and then the $\chi^2$, with and without angular restriction, for the original potentials (columns 3 and 4) and for the energy-dependent potentials resulting from our fit (columns 5 and 6).}
\label{tab:chi2}
\end{table}

Starting from  PB, we fit each data set, by allowing both the real and imaginary depths to vary ($V_v$ and $W_s$), while keeping the rest of the original parameterization.  While the resulting $V_v$ was very close to the original potential, the data required a significant variation of $W_s$ with energy (mostly linear) and with the target. We repeated this procedure using the TPM potential as starting point, varying the depth of the real part $V_v$, and both volume and surface depths for the imaginary part ($W_v$ and $W_s$). We found again that $V_v$ reflected weak energy and target dependencies, but there were strong variations of $W_v$ and $W_s$. This preliminary study \cite{bacq-masters} pointed toward the need for an explicit energy dependence in the imaginary part of the optical potential.

Considering only the effect of antisymmetrization, one might expect a Gaussian non-locality as introduced in these phenomenological potentials, with a range roughly of the size of the nucleon. However there is no reason to expect this shape and range should account for the complex scattering dynamics \cite{Fraser_epja2008, Rawitscher_prc1994, Rawitscher_npa1987}. Our hypothesis is that channel-coupling effects would primarily affect the  absorptive term. In addition to the energy dependence, it should also be  target dependent. We now turn to local phenomenological potentials to obtain insight. \\

\begin{table}[t]
\centering
\begin{tabular}{|  c  | c | c |}
\hline
 & PB-E & TPM-E \\
 \hline
$a$ & $-0.017 \pm 0.015$ & $0.20 \pm 0.0040$ \\
$b$ (MeV) & $0.74 \pm 0.46$ & $4.5 \pm 0.5$ \\
$c$ (MeV) &  $11.94 \pm 0.38$ & $12.15 \pm 0.40$ \\
$d$ & $0.34 \pm 0.011$ & $0.018 \pm 0.0085$ \\
$e$ (MeV) &  $-2.00 \pm 0.25$ & $0.36 \pm 0.26$  \\ 
\hline
\end{tabular}
\caption{Best fit parameters using PB or TPM as starting points for the minimization, fitting data with $\theta<100^\circ$. }
\label{tab:par}
\end{table}


{\it Energy and asymmetry dependences:}
A strong energy dependence in the depth of the optical potential is usually obtained when extracting local global optical potentials \cite{koning_np2003,ch89,bg69}.
A simple parameterization is provided in Becchetti and Greenlees (BG) \cite{bg69}, where the global optical potential was derived for targets with mass $A>40$  and at energies $E<50$ MeV. The  corresponding parameterizations for $V_v$, $W_v$, and $W_s$ all contain a linear dependence
in energy and/or asymmetry.
While we do not expect this potential to compare directly to the non-local potentials under study here, we borrow the simple form of the isotopic dependence of Ref.~\cite{bg69} in our analysis.

Since our independent fits show no need for an energy-dependent real part in the optical potential, we keep it constant.
We parametrize the imaginary depths as
\begin{equation}
W_s=a\,E +b\,(N-Z)/A  +c, \;\;\;\;\; W_v=d\,E + e  \;. 
\end{equation}
We then fit these five parameters to all 24 sets of elastic data.  During this fit, all other parameters are kept at their original value, and the geometry for the imaginary volume term in the PB potential is taken to be the same as the imaginary volume term from TPM.
We do this for the PB-like (PB-E) and TPM-like (TPM-E) potentials. We take only data up to $\theta=100^\circ$, to avoid distortions of the potential in regions where the optical model may not be reliable.
We estimate the uncertainties on the parameters from the covariance matrix.
The resulting parameterizations are given in Table~\ref{tab:par}.

\begin{figure}[t]
\center
\includegraphics[scale=0.33]{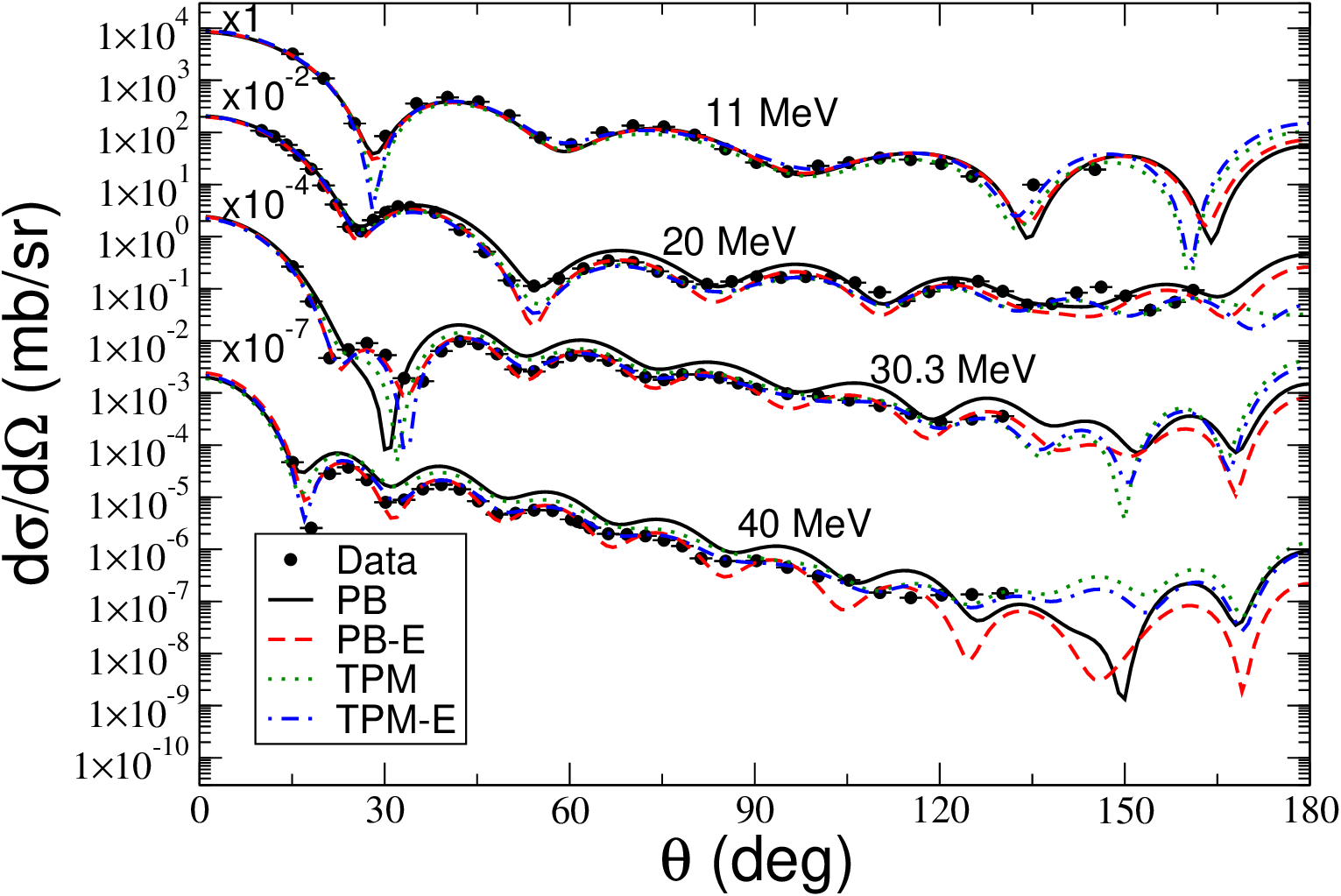}
\caption{(Color online) Angular distributions for $^{208}$Pb$(n,n)$$^{208}$Pb at $E=11$, 20, 30.3, 40 MeV.  Data is from \cite{208Pb791126data,208Pb2022data,208Pb3040data}.}
\label{fig-pb}
\end{figure}

\begin{figure}[t]
\center
\includegraphics[scale=0.33]{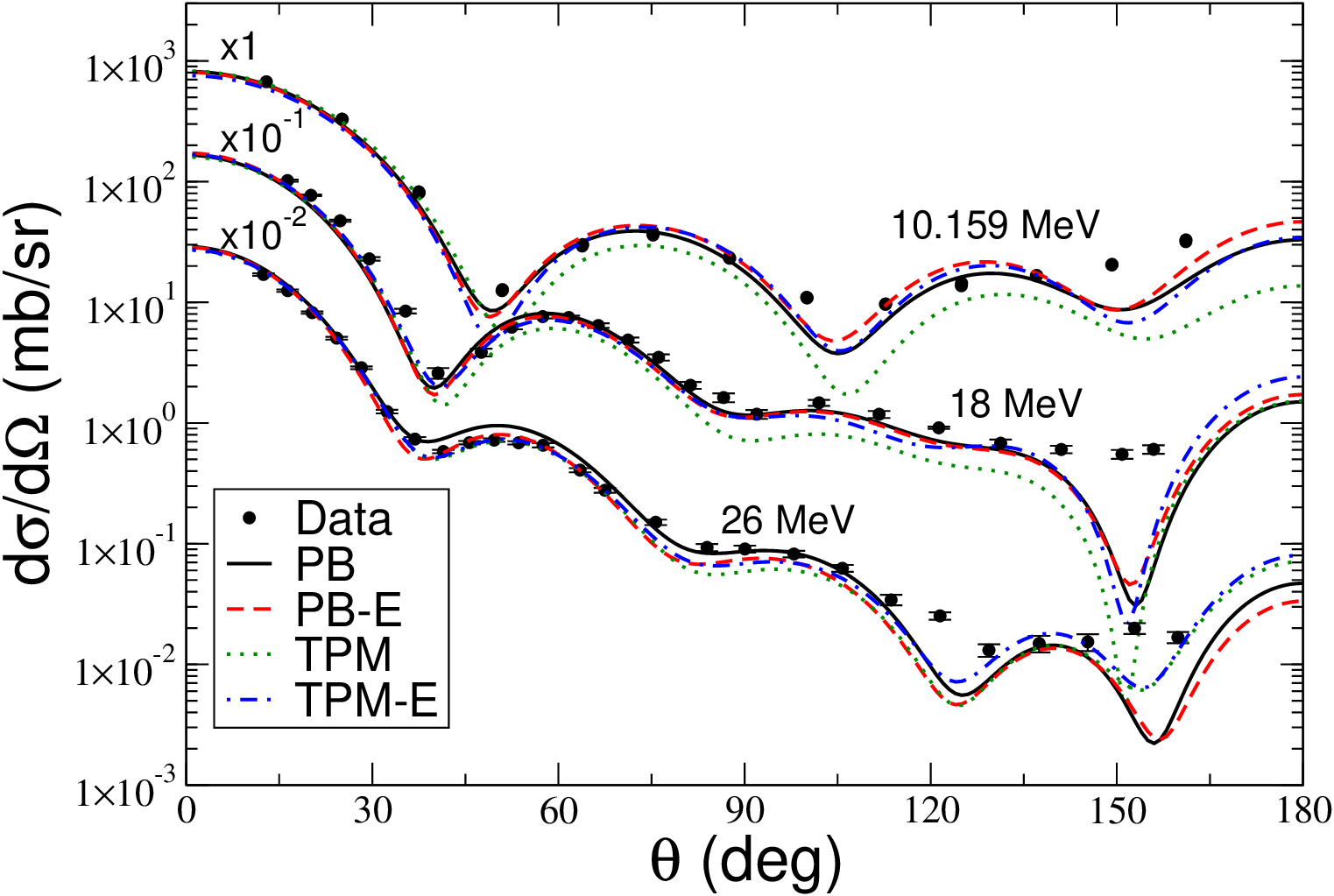}
\caption{(Color online) Angular distributions for $^{27}$Al$(n,n)$$^{27}$Al at $E=10.159$, 18, 26 MeV.  Data is from \cite{27Al10data,27Al1826data}.}
\label{fig-al}
\end{figure}

\begin{figure}[t]
\center
\includegraphics[scale=0.33]{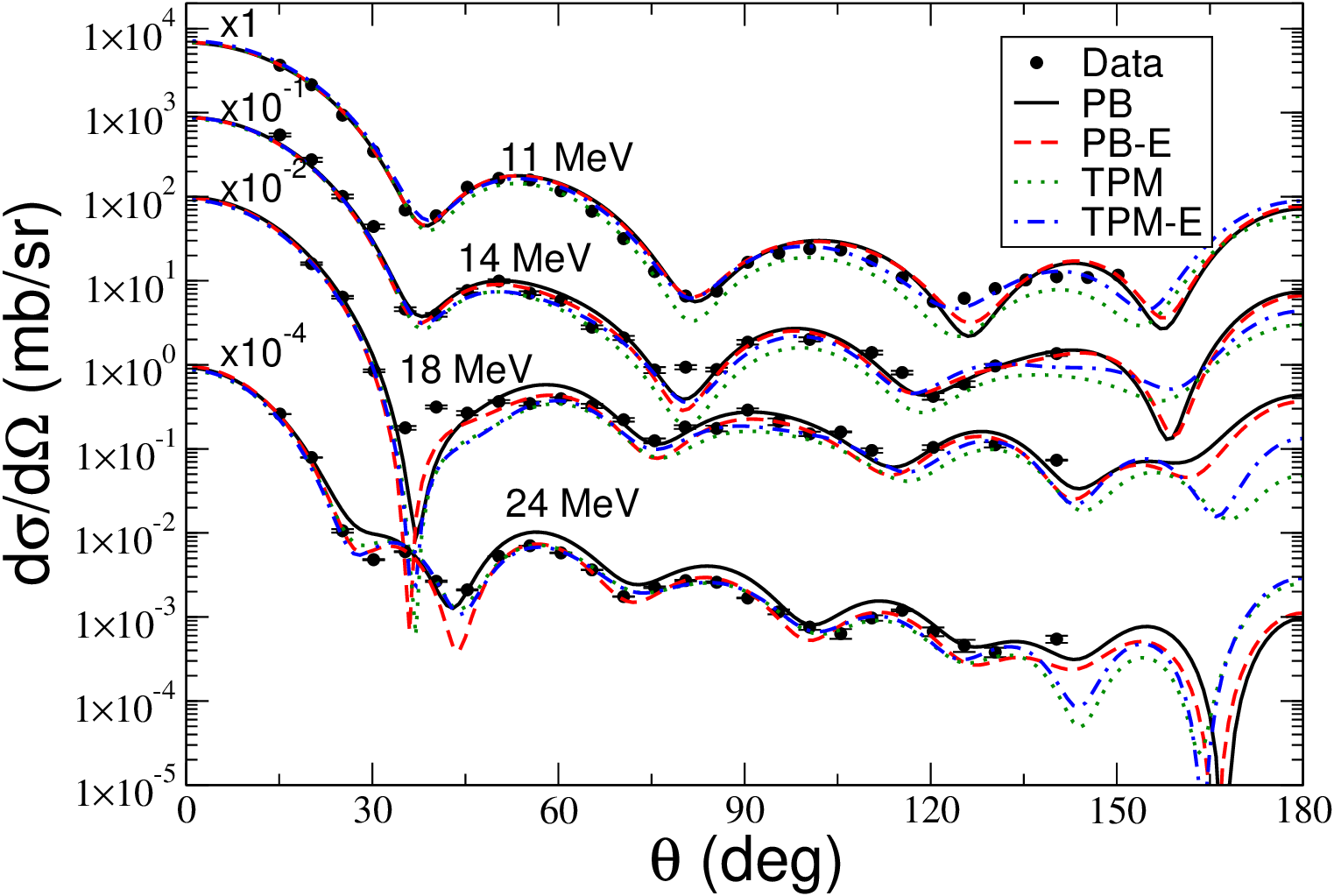}
\caption{(Color online) Angular distributions for $^{118}$Sn$(n,n)$$^{118}$Sn at $E=11$, 14, 18, 24 MeV.  Data is from \cite{118Sn1124data,118Sn14data}.}
\label{fig-sn}
\end{figure}

We compare the $\chi^2$ for  these two parameterizations with those of the original potentials.
We summarize the results for the $\chi^2$ in columns 5 and 6 of Table~\ref{tab:chi2}. 
No matter whether you consider the lower or higher energy bin, all angles or the restricted angular region, the energy dependence always provides a significant improvement in the description of the data.
This improvement can be very large for those cases in which the original potential performs poorly.

We also present in Fig. \ref{fig-pb} the angular distributions for neutron elastic scattering on $^{208}$Pb at four different energies. The results for elastic scattering using the original PB (solid black) and TPM (dotted green) potentials are compared with those obtained using our energy-dependent parameterizations PB-E (dashed red) and TPM-E (dot-dashed blue), and with the data of Refs.~\cite{208Pb791126data,208Pb2022data,208Pb3040data}. Cross sections are multiplied by arbitrary factors for readability. 
The original PB potential describes the angular distribution well for the lowest energies, while the original TPM potential does an excellent job at $E=30$ MeV. 
As mentioned earlier, both PB and TPM provide cross sections that are significantly far from the data at 40 MeV.
Both the PB-E and TPM-E parameterizations are effective in describing the data across the whole energy range. 

The two sets of parameters PB-E and TPM-E, shown in Table II,  are very different. The PB-E potential is consistent with no energy dependence in the surface term of its imaginary part and a robust energy dependence in the volume term (the predicted slope $d$ is larger than the associated error by an order of magnitude). On the other hand, the TPM-E parameterization has a strong energy dependence in the surface term and a weaker -- but non-zero -- energy dependence in the volume term. This set of data hence does not constrain the details of $W_s$ and $W_v$, but regardless of the results, energy dependence is required.
We do not expect the energy dependence of the non-local optical potential to be identical to the local counterpart.
Nevertheless, it is interesting to note that the slopes $a$ and $d$ we obtain in the non-local parameterizations are of the same order of magnitude as in the local BG potential \cite{bg69}.

Interestingly, the energy dependence also relies on the geometry of the potential and parameter restrictions imposed during the fit.  Table \ref{tab:PBfix} shows two alternate parameterizations that were obtained starting from PB.  Because PB does not contain an imaginary volume term, the choice of geometry for this term is rather arbitrary.  If instead of using the TPM geometry, we choose the geometry of the imaginary surface term in PB (many local potentials impose that these two terms have the same geometry), and perform the fit for angles less than $100^\circ$, we find the parameters given in column two of Table \ref{tab:PBfix} (PB-PB).  Comparing these to the PB-E parameterization, we now find a robust (non-zero) energy dependence in both of the imaginary terms, with a similar asymmetry term.  However, $e$ is more negative leading to a larger range of energies for which $W_v$ would be negative (defined by $E<-e/d$).

\begin{table}[t]
\centering
\begin{tabular}{|  c  | c | c |}
\hline
 & PB-PB & PB-00 \\
 \hline
$a$ & $0.017 \pm 0.0044$ & $0.14 \pm 0.0076$ \\
$b$ (MeV) & $1.3 \pm 0.59$ & $0.066 \pm 0.57$ \\
$c$ (MeV) &  $12.60 \pm 0.097$ & $9.3 \pm 0.15$ \\
$d$ & $0.31 \pm 0.0056$ & $0.23 \pm 0.065$ \\
$e$ (MeV) &  $-2.33 \pm 0.068$ & $9.6\times 10^{-5} \pm 5.4 \times 10^{-3}$  \\ 
\hline
\end{tabular}
\caption{Best fit parameterization starting with PB assuming the same geometry for the two imaginary terms (column 2) and imposing $e\ge0$ (column 3), fitting data with $\theta<100^\circ$.}
\label{tab:PBfix}
\end{table}

In addition, we can impose the restriction $e\ge0$ when we perform the fit using the PB geometry. The resulting parameters are given in column 3 of Table \ref{tab:PBfix} (PB-00).  In this case, $e$ is consistent with zero, meaning that this is probably not a true minimum of the system but rather an artifact of hitting one of the imposed bounds.  Still, we use this to illustrate the differences that can arise due to different restrictions that are imposed.  In this case, there is still a strong energy dependence in both imaginary terms, but we also find an asymmetry that is undetermined within its associated error.  While the need for an energy dependence in the imaginary terms is robust, the specific parameters are heavily influenced by the constraints that are included in the fitting.  This implies that this set of elastic scattering data is not enough to constrain the details of the imaginary potential.

The asymmetry dependence obtained in both parameterizations of Table \ref{tab:par} are much smaller than the BG one: PB-E predicts a weak $b<1$ MeV, while TPM-E predicts a slightly larger  $b =4.5$ MeV.
Moreover, we find no need for an asymmetry dependence in the real part of the potential, contrary to Becchetti and Greenlees.
Given the large difference in $b$ obtained in both parameterizations and the significant corresponding uncertainties, we do not think the resulting asymmetry dependence is robust. For a meaningful study of the asymmetry dependence, one will have to include both neutron and proton data, and potentially charge-exchange measurements, which is beyond the scope of the current study.
\\


{\it Predictions:} 
Finally we use the PB-E and TPM-E parameterizations to make predictions for a few cases that are not included in our fit. We choose spherical nuclei in very different parts of the nuclear chart, namely $^{27}$Al and $^{118}$Sn, for which a wealth of neutron elastic-scattering data exists.
In Figs. \ref{fig-al} and \ref{fig-sn} we show the neutron elastic-scattering cross sections on $^{27}$Al and $^{118}$Sn, respectively.
The calculations with the original potentials and our energy-dependent parameterizations are compared to data \cite{27Al10data,27Al1826data,118Sn1124data,118Sn14data}. 
In both cases, the energy dependent parameterizations provide good predictions for the angular distributions over the whole energy range, which indicates that the interpolation over asymmetry is valid.
Here also the original TPM potential does poorly at the lowest energies, while the opposite is true for the original PB potential.\\


{\it Summary and Outlook:}
We have investigated the parameterization of non-local optical potentials that describe the elastic scattering of neutrons off nuclear targets.
We have found that neither the PB \cite{Perey_np1962} nor the TPM \cite{Tian_ijmpe2015} potentials are able to provide a good description of elastic-scattering data across the energy range $E=5-40$ MeV.
To correct this, we have developed two new parameterizations by including in the original non-local PB and TPM potentials an energy dependence inspired from the BG local optical potential \cite{bg69}.
Our results for $^{40}$Ca, $^{90}$Zr and $^{208}$Pb, the nuclei included in the fit, demonstrate a clear improvement in the description of the angular distributions, with the $\chi^2$ improving by factors of 2, 5, or even 10 for specific data sets. These energy dependent fits, PB-E and TPM-E, are also able to make predictions for nuclei not included in the fit. 

While  we find that some details of the PB-E and TPM-E parameterizations are not robust or unique, like their asymmetry dependence, 
our study clearly shows that, when including a standard Gaussian nonlocality in the optical potential, one still needs a significant energy dependence. This energy dependence  can be parameterized with a simple linear term. 
The need for energy dependence in addition to nonlocality is in agreement with the expectation from microscopic theories but it goes against the belief by a significant fraction of the nuclear-reaction community that non-locality alone can remove the energy dependence of optical potentials. Given that there is an interplay between nonlocality and energy-dependence, we must note that our conclusions are specific to our choice of a standard Gaussian nonlocality.
This study calls for the development of a new global non-local energy-dependent optical potential, encompassing a larger range of data and a more varied array of observables. One important next step for this work on neutron scattering is to allow for a non-Gaussian nonlocality provided by ab initio calculations. Also important is to  analyse proton elastic data on similar systems and  explore long isotopic chains to reliably constrain the asymmetry term.

\begin{acknowledgments}
This work was supported by the National Science
Foundation under Grants No. PHY-1068571 and PHY-1403906, the Stewardship Science Graduate Fellowship program under Grant No. DE-NA0002135, and the
Department of Energy under Contract No. DE-FG52-
08NA28552.
P.~C. acknowledges the support of the Deutsche Forschungsgesellschaft (DFG) with the Collaborative Research Center 1245.
This work was supported in part by the European Union's Horizon 2020 research and innovation program under Grant Agreement No. 654002.
All neutron-scattering data were collected on the \href{https://www-nds.iaea.org/exfor/exfor.htm}{EXFOR} database.
\end{acknowledgments}

\bibliography{nonlocal}

\end{document}